\documentclass[aps,prl,twocolumn,groupedaddress,showpacs,amsmath,amssymb,floatfi
x,superscriptaddress]{revtex4}
\usepackage{graphicx}
\usepackage{times}

\newcommand{\nuebar}{$\overline{\nu}_{e}$}

\begin{document}

\title{Search for the Invisible Decay of Neutrons with KamLAND}

\newcommand{\tohoku}{\affiliation{Research Center for Neutrino
    Science, Tohoku University, Sendai 980-8578, Japan}}
\newcommand{\alabama}{\affiliation{Department of Physics and
    Astronomy, University of Alabama, Tuscaloosa, Alabama 35487, USA}}
\newcommand{\lbl}{\affiliation{Physics Department, University of
    California at Berkeley and \\ Lawrence Berkeley National Laboratory, 
Berkeley, California 94720, USA}}
\newcommand{\caltech}{\affiliation{W.~K.~Kellogg Radiation Laboratory,
    California Institute of Technology, Pasadena, California 91125, USA}}
\newcommand{\drexel}{\affiliation{Physics Department, Drexel
    University, Philadelphia, Pennsylvania 19104, USA}}
\newcommand{\hawaii}{\affiliation{Department of Physics and Astronomy,
    University of Hawaii at Manoa, Honolulu, Hawaii 96822, USA}}
\newcommand{\kansas}{\affiliation{Department of Physics,
    Kansas State University, Manhattan, Kansas 66506, USA}}
\newcommand{\lsu}{\affiliation{Department of Physics and Astronomy,
    Louisiana State University, Baton Rouge, Louisiana 70803, USA}}
\newcommand{\unm}{\affiliation{Physics Department, University of New
    Mexico, Albuquerque, New Mexico 87131, USA}}
\newcommand{\stanford}{\affiliation{Physics Department, Stanford
    University, Stanford, California 94305, USA}}
\newcommand{\ut}{\affiliation{Department of Physics and
    Astronomy, University of Tennessee, Knoxville, Tennessee 37996, USA}}
\newcommand{\tunl}{\affiliation{Triangle Universities Nuclear
    Laboratory, Durham, North Carolina 27708, USA and \\
Physics Departments at Duke University, North Carolina State University,
and the University of North Carolina at Chapel Hill}}
\newcommand{\ihep}{\affiliation{Institute of High Energy Physics,
    Beijing 100039, People's Republic of China}}
\newcommand{\cnrs}{\affiliation{CEN Bordeaux-Gradignan, IN2P3-CNRS and
    University Bordeaux I, F-33175 Gradignan Cedex, France}}

\newcommand{\aticrrnow}{\altaffiliation{Present address: ICRR, 
    University of Tokyo, Gifu, Japan}}
\newcommand{\aticeppnow}{\altaffiliation{Present address: ICEPP,
    University of Tokyo, Tokyo, Japan}}
\newcommand{\atimperialnow}{\altaffiliation{Present address: Imperial
    College London, UK}}
\newcommand{\atlanlnow}{\altaffiliation{Present address: LANL, Los
    Alamos, NM 87545, USA}}
\newcommand{\atiasnow}{\altaffiliation{Present address: School of
    Natural Sciences, Institute for Advanced Study, Princeton, NJ
    08540, USA}}
\newcommand{\atksunow}{\altaffiliation{Present address: KSU, Manhattan, KS 66506, 
USA}}
\newcommand{\atdubnanow}{\altaffiliation{Present address: DLNP, JINR, Dubna, Russia}}

\author{T.~Araki}\tohoku
\author{S.~Enomoto}\tohoku
\author{K.~Furuno}\tohoku
\author{Y.~Gando}\tohoku
\author{K.~Ichimura}\tohoku
\author{H.~Ikeda}\tohoku
\author{K.~Inoue}\tohoku
\author{Y.~Kishimoto}\tohoku
\author{M.~Koga}\tohoku
\author{Y.~Koseki}\tohoku
\author{T.~Maeda}\tohoku
\author{T.~Mitsui}\tohoku
\author{M.~Motoki}\tohoku
\author{K.~Nakajima}\tohoku
\author{K.~Nakamura}\tohoku
\author{H.~Ogawa}\tohoku
\author{M.~Ogawa}\tohoku
\author{K.~Owada}\tohoku
\author{J.-S.~Ricol}\tohoku
\author{I.~Shimizu}\tohoku
\author{J.~Shirai}\tohoku
\author{F.~Suekane}\tohoku
\author{A.~Suzuki}\tohoku
\author{K.~Tada}\tohoku
\author{S.~Takeuchi}\tohoku
\author{K.~Tamae}\tohoku
\author{Y.~Tsuda}\tohoku
\author{H.~Watanabe}\tohoku
\author{J.~Busenitz}\alabama
\author{T.~Classen}\alabama
\author{Z.~Djurcic}\alabama
\author{G.~Keefer}\alabama
\author{D.S.~Leonard}\alabama
\author{A.~Piepke}\alabama
\author{E.~Yakushev}\atdubnanow\alabama
\author{B.E.~Berger}\lbl
\author{Y.D.~Chan}\lbl
\author{M.P.~Decowski}\lbl
\author{D.A.~Dwyer}\lbl
\author{S.J.~Freedman}\lbl
\author{B.K.~Fujikawa}\lbl
\author{J.~Goldman}\lbl
\author{F.~Gray}\lbl
\author{K.M.~Heeger}\lbl
\author{L.~Hsu}\lbl
\author{K.T.~Lesko}\lbl
\author{K.-B.~Luk}\lbl
\author{H.~Murayama}\lbl
\author{T.~O'Donnell}\lbl
\author{A.W.P.~Poon}\lbl
\author{H.M.~Steiner}\lbl
\author{L.A.~Winslow}\lbl
\author{C.~Jillings}\caltech
\author{C.~Mauger}\caltech
\author{R.D.~McKeown}\caltech
\author{P.~Vogel}\caltech
\author{C.~Zhang}\caltech
\author{C.E.~Lane}\drexel
\author{T.~Miletic}\drexel
\author{G.~Guillian}\hawaii
\author{J.G.~Learned}\hawaii
\author{J.~Maricic}\hawaii
\author{S.~Matsuno}\hawaii
\author{S.~Pakvasa}\hawaii
\author{G.A.~Horton-Smith}\kansas
\author{S.~Dazeley}\lsu
\author{S.~Hatakeyama}\lsu
\author{A.~Rojas}\lsu
\author{R.~Svoboda}\lsu
\author{B.D.~Dieterle}\unm
\author{J.~Detwiler}\stanford
\author{G.~Gratta}\stanford
\author{K.~Ishii}\stanford
\author{N.~Tolich}\stanford
\author{Y.~Uchida}\stanford
\author{M.~Batygov}\ut
\author{W.~Bugg}\ut
\author{Y.~Efremenko}\ut
\author{Y.~Kamyshkov}\ut
\author{A.~Kozlov}\ut
\author{Y.~Nakamura}\ut
\author{H.J.~Karwowski}\tunl
\author{D.M.~Markoff}\tunl
\author{R.M.~Rohm}\tunl
\author{W.~Tornow}\tunl
\author{R.~Wendell}\tunl
\author{M.-J.~Chen}\ihep
\author{Y.-F.~Wang}\ihep
\author{F.~Piquemal}\cnrs

\collaboration{The KamLAND Collaboration}\noaffiliation

\date{\today}

\begin{abstract}

The Kamioka Liquid scintillator Anti-Neutrino Detector (KamLAND) is used in a 
search for single neutron or two neutron intra-nuclear  
disappearance that would produce holes in the $\it{s}$-shell energy level 
of $^{12}$C nuclei. Such holes could be created as a result of nucleon 
decay into invisible modes ($inv$), e.g. $n \rightarrow 3\nu$ or $nn \rightarrow 2\nu$. 
The de-excitation of the corresponding daughter nucleus results in a sequence of space 
and time correlated events observable in the liquid scintillator detector. 
We report on new limits for one- and two-neutron disappearance:  
$\tau(n\rightarrow inv)> 5.8\times 10^{29}$ years and 
$\tau (nn \rightarrow inv)> 1.4 \times 10^{30}$ years at 90\% CL. These results
represent an improvement of factors of $\sim$3 and $>10^4$ over previous experiments.

\end{abstract}

\pacs{13.30.Ce, 11.30.Fs, 14.20.Dh, 29.40.Mc}

\maketitle
Baryon number violation is an important signal of physics
beyond the Standard Model. Grand unified
theories suggest that processes such as $p \rightarrow e^+ \pi^0$ and
$n \rightarrow K^0 \bar{\nu}$ are important but suppressed by powers 
of the grand unification scale. However, 
baryon number violation is not limited to grand unified theories.  In
fact, baryon number is an ``accidental'' symmetry arising from the pattern of
particles in, and renormalizability of, the Standard
Model. New particles or new physics such as supersymmetry or
extra dimensions might also be the cause of baryon
number violation. Moreover, the suppression may only be at the TeV scale, instead of the much higher
grand-unified scale.

The Particle Data Group \cite{PDG} lists more than 70 possible
modes of nucleon decay which conserve 
electric charge, energy-momentum, and angular momentum. Experimental
lifetime limits of 10$^{30}$ years for all but a few of these
modes have been obtained. The decay modes with the poorest limits are
so called ``invisible'' modes, such as $n \rightarrow \nu$'s and $nn \rightarrow
\nu$'s.  Invisible modes are dominant in some models~\cite{Mohapatra:2002ug}.

Invisible decay modes involving bound neutrons are
potentially detectable even though no energetic charged particles are
produced directly in the decay. If we assume that the unobserved decay products
carry away most of the neutron rest-mass energy, then the residual nucleus, 
$_\text{\;\;\,Z}^\text{A-1}$X or
$^\text{A-2}_\text{\;\;\,Z}$X, has a hole or holes in the
previously occupied shell. Particles emitted in the nuclear de-excitation 
of the daughter are the experimental signature for the process. 
Detectability is independent of the specifics of the process as long as the
rest-mass energy is carried away by the undetected particles.

The SNO Collaboration established the best single neutron disappearance limit of
$\tau(n\rightarrow inv)> 1.9 \times 10^{29}$ years (90\%~CL) by searching for the de-excitation $\gamma$ rays
following neutron disappearance in $^{16}$O~\cite{Ahmed}.
The best limit for two-neutron disappearance, $\tau (nn \rightarrow inv) 
> 4.9 \times 10^{25}$~y (90\%~CL), was set by the Borexino Collaboration~\cite{Back} 
by searching for possible decays of unstable nuclides 
resulting from $nn$ disappearance in $^{12}$C, $^{13}$C, and $^{16}$O. 

KamLAND, currently the world's largest low-background
liquid scintillator (LS) detector (the LS composition is CH$_{1.97}$), has a low energy threshold 
($<$ 1\,MeV) and good energy and spatial resolution, making it 
well suited for neutron disappearance searches. The LS is 
contained in a 13-m-diameter transparent nylon-based balloon suspended in non-scintillating 
purified mineral oil. The scintillator is viewed by 1325 17-inch Hamamatsu 
and 554 slower 20-inch Hamamatsu photomultiplier tubes (PMTs) mounted on 
the inner surface of an 18-m-diameter spherical stainless-steel tank. The 
inactive mineral oil buffer serves as passive shielding 
against external backgrounds such as $^{208}$Tl $\gamma$ rays coming from the 
PMT's glass and nearby rocks. 
A 3200\,\,metric ton water-Cherenkov detector surrounds the stainless-steel sphere,
acting as a cosmic-ray muon veto. The muon rate in the central detector is 0.34\,Hz. 
Details of KamLAND can be found elsewhere~\cite{Eguchi}.

In this study, we consider the disappearance of neutrons only from 
the fully occupied $\it{s}$-shell in $^{12}\text{C}$ leaving the 
daughter nucleus in a highly excited state. This results in subsequent de-excitation 
with the emission of secondary particles ($p, n, d, \alpha, \gamma$) and the 
possible further decay of the residual radioactive nuclei. The probabilities for 
the various de-excitation modes following one- and two-neutron disappearance in 
$^{12}\text{C}$ were estimated with a statistical nuclear model by Kamyshkov and 
Kolbe \cite{Kolbe}. Some of these modes, producing a sequence of time-and-space-correlated
events, are detectable in KamLAND. Here we report on a search for the
four decay modes listed in Table~\ref{tab1}, each producing a correlated triple signal in the detector. 

\begin{table}[t]
\caption{ \label{tab1}Branching ratios for the $^{11}$C$^*$ ($^{10}$C$^*$) 
\cite{Kolbe} de-excitation modes after neutron (two neutron) disappearance from 
the $s_{1/2}$-state in $^{12}$C and the experimental signature (number of 
time-correlated hits) for observation of these modes in KamLAND.}
 \begin{center}   \begin{tabular}{l|l|c|c} 
\hline \hline
     Decay & Daughter                          & Branching  & Exp.   \\
     mode  & (decay, T$_{1/2}$, and Q$_{EC}$)  & ratio    & sign.  \\
           &                                   & \%      & (hits) \\ 
\hline\hline
    (n$1$) $^{11}$C(n)               & $^{10}$C$_{\rm gs}$($\beta^+$; 19.3 s, 3.65 
MeV)      &  3.0  &  3 \\
    (n$2$) $^{11}$C(n,$\gamma$)      & $^{10}$C$_{\rm gs}$($\beta^+$; 19.3 s, 3.65 
MeV)   &  2.8  &  3 \\ 
\hline
    (nn$1$) $^{10}$C(n)               & $^{9}$C($\beta^{+}$, 0.127 s, 16.5 MeV)
       &  6.2  & 3 \\
    (nn$2$) $^{10}$C(n,p)             & $^{8}$B($\beta^{+}\alpha$, 0.77 s, 18 MeV)
       &  6.0  & 3 \\
\hline \hline
   \end{tabular}
  \end{center}
\end{table} 

According to Ref.~\cite{Kolbe}, neutrons produced in these modes 
have energies up to 45\,MeV. In the LS, the first scintillation signal 
arises from neutron-proton elastic 
scattering, 4.4\,MeV $\gamma$~rays from the inelastic neutron 
scattering from $^{12}$C, and detectable secondaries of other neutron 
interactions in the scintillator. Neutrons 
thermalize and capture, mostly on protons ($\tau_{\text{capture}}$~$\sim210\,\mu s$), 
producing a second 2.2\,MeV $\gamma$~ray signal.  A third correlated signal 
comes from the $\beta^{+}$~decay of the residual nuclei ($^{10}$C, $^{9}$C or 
$^{8}$B). The correlated triple signal is a powerful tool for identifying these decays
in the presence of backgrounds.

The limits presented here are based on data collected between March 5, 
2002 and October 31, 2004. Initially, KamLAND had 17-inch PMTs only operational with 
22\% photo-cathode coverage and 7.3\%$/\sqrt{E(\text{MeV})}$ 
energy resolution. After February 27, 2003 the photo-cathode coverage increased to 34\% by including 20-inch PMTs, which 
improved the energy resolution to 6.2\%$/\sqrt{E(\text{MeV})}$. The 
location of the event in the detector is reconstructed from the arrival times
measured with the PMTs. Typical spatial 
resolution for the reconstruction procedure at the detector center, for a point-like 
energy deposition, is $\sim 20.6\,\text{cm}$~($\sim 
24.0\,\text{cm}$)/$\sqrt{\text{E(MeV})}$ with 20-inch (without 20-inch)
PMT data.  The energy of the event is reconstructed using the number of 
detected photo-electrons after correcting for position and PMT gain variations. 
The response of the detector is studied and monitored with periodic deployments of $\gamma$ ray 
($^{203}$Hg, $^{68}$Ge, $^{65}$Zn, and $^{60}$Co) and neutron sources (Am-Be) 
along the central vertical axis of the detector. Natural radio-isotopes remaining in the
detector materials such as $^{40}$K and bismuth-polonium from U/Th chains are also used for calibrations.
\begin{table}[t]
\caption{ \label{tab2} Top section: Selection criteria used in the neutron (two 
neutron) disappearance search. Bottom section: detection efficiency $\epsilon$ for the decay modes $\text{n}1$, $\text{n}2$, 
$\text{nn}1$ and $\text{nn}2$ given in Table~\ref{tab1}.}

 \begin{center}
   \begin{tabular}{l|c|c} \hline \hline
      Quantity &\;\; $n$ disappearance\;\; &\;\; $nn$ disappearance\;\;   
       \\ \hline\hline
   \; $\text{R}_{1,2,3}$\;\;\;\;\;\; [m]\; & 5.0 & 5.5\\
   \; $\text{R}_{XY3}$\;\;\;\;\;\; [m]\; & $>$1.0 & $>$1.0\\
    \hline
   \; $\Delta \text{R}_{12}$\;\;\;\;\;\; [m]\; & 2.0 & 2.0 \\
   \; $\Delta \text{R}_{13}$\;\;\;\;\;\; [m]\; & 0.8 & 1.0 \\
    \hline
   \; $\Delta \text{T}_{12}$\;\;\;\;\;\; [$\mu$s]\; & 0.5\,-\,1000 & 0.5\,-\,1000 \\
   \; $\Delta \text{T}_{13}$\;\;\;\;\;\; [s]\; & 0.003\,-\,70 & 0.003\,-\,6 \\
    \hline
   \; $\text{E}_{1}$\;\;\;\;\;\; [MeV]\; & 0.9\,-\,25 &0.9\,-\,40 \\
   \; $\text{E}_{2}$\;\;\;\;\;\; [MeV]\; & 1.8\,-\,2.6 & 1.8\,-\,2.6\\
   \; $\text{E}_{3}$\;\;\;\;\;\; [MeV]\; & 1.5\,-\,3.8 & 3.1\,-\,18.0 \\
   \hline \hline
   \; $\epsilon_{\text{n}1 (\text{nn}1)}$ &0.430\,$\pm$\,0.027 &0.680\,$\pm$\,0.032 \\
   \; $\epsilon_{\text{n}2 (\text{nn}2)}$ &0.651\,$\pm$\,0.033 &0.678\,$\pm$\,0.032 \\
   \hline \hline
   \end{tabular}
  \end{center}
\end{table}
Because of the external $\gamma$ ray background, a radial 
fiducial volume cut of 5\,m (5.5\,m) for the $n$ ($nn$) disappearance search is
applied. This corresponds to a fiducial mass of 408.5 (543.7)\, metric tons, corresponding to 3.48$\times$10$^ {\text 31}$  
(4.63$\times$10$^{\text 31}$) $s$-shell neutrons in $^{12}$C nuclei. The total 
livetime, after correction for muon-associated cuts to avoid backgrounds due to muon-spallation events, is 749.8 days 
(751.4\,days) for the 5.0\,m (5.5\,m) fiducial volume. 

The correlated triple event-selection cuts are summarized in Table \ref{tab2}. 
The indices refer to the first, second and third energy deposits (hits) in each candidate event. R is the radial 
position of the hit, $\text{R}_{XY}$ is the distance to the vertical central axis, E is the reconstructed energy of the hit, 
$\Delta$R and $\Delta$T are the position and time differences between the hits. 
After applying all cuts to the dataset, the number of observed $n$ ($nn$) disappearance candidate events is 1 (0), 
compatible with the expected background, as discussed next.

\begin{figure}[t]
\includegraphics[scale=0.43]{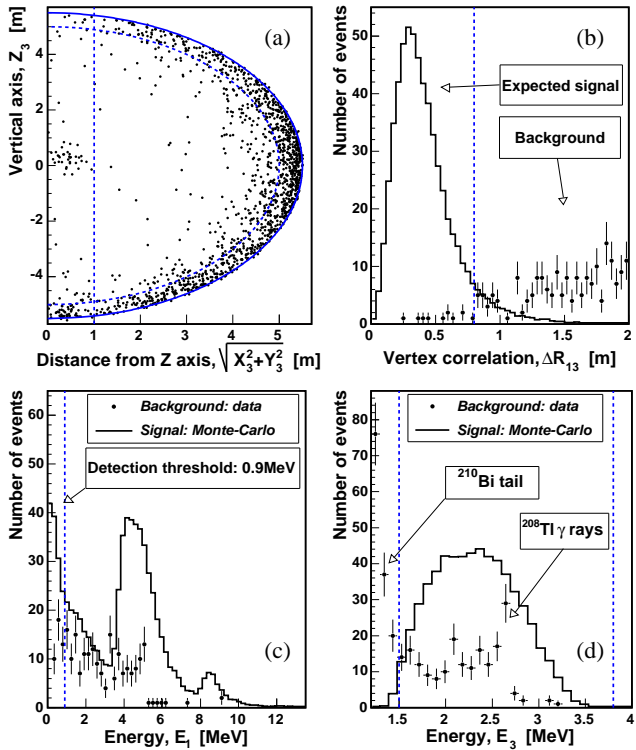}
\put(-145,272){(a)}
\put(-27,272){(b)}
\put(-145,40){(c)}
\put(-27,40){(d)}
\caption{$n$ disappearance search: 
(a) correlated triple background events in the 150\,s-1000\,s off-time window for the 
3$^{rd}$ hit inside the 5.5\,m detector radius; 
(b-d) events inside the 5\,m fiducial volume and outside of the 1\,m cylinder cut around the central axis 
are compared to the expected signal for $\tau$\,=\,4$\times$10$^{\text{27}}$\,y, generated with the 
Monte-Carlo.
Dashed lines indicate cuts used in the data analysis.}
\label{fig:ndis}
\end{figure} 

\begin{figure}[t]
\includegraphics[scale=0.43]{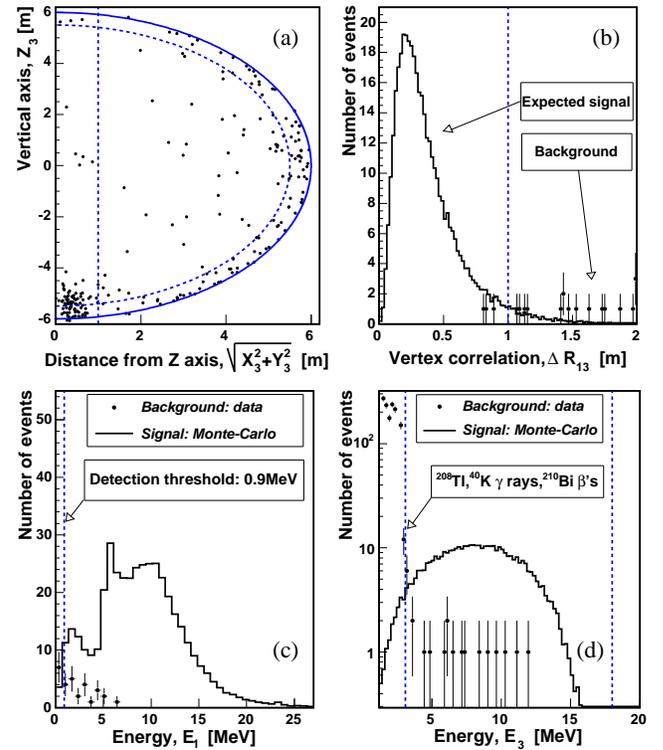}
\put(-145,272){(a)}
\put(-24,272){(b)}
\put(-145,40){(c)}
\put(-30,40){(d)}
\caption{$nn$ disappearance search:
(a) correlated triple background events in the 10\,s-1000\,s off-time window for the 
3$^{rd}$ hit inside 6\,m detector radius; 
(b-d) events inside the 5.5\,m fiducial volume and outside of the 1\,m cylinder cut around the central axis 
are  compared to the expected signal for $\tau$\,=\,10$^{\text{28}}$\,y, generated with the Monte Carlo.
Dashed lines indicate cuts used in the data analysis.}
\label{fig:nndis}
\end{figure} 
No other process is expected to produce triple events
with sufficient rates to result in backgrounds for the present search.
Backgrounds to the $n$ and $nn$ disappearance search originate from doubly correlated signatures, 
such as those derived from reactor \nuebar\ events or other processes such as neutrons 
produced in $^{13}$C($\alpha$,n)$^{16}$O followed by an uncorrelated coincidence within the selection cuts. The third uncorrelated hit
is dominated by the tail of the $^{210}$Bi $\beta$-spectrum and $^{208}$Tl $\gamma$~rays (see 
Fig.~\ref{fig:ndis}d and Fig.~\ref{fig:nndis}d). The data analysis cuts were optimized
to suppress the acidental coincidence background and to improve the signal/background ratio. 
For the case of the $n$ disappearance search, Fig.~\ref{fig:ndis}a shows the location of the accidental
background events within the 5.5\,m fiducial volume using a relaxed $\Delta$R$_{13}$\,$<$\,2\,m cut.
Events inside the 5\,m fiducial volume and outside of the 1\,m cylinder cut around the central axis 
(to remove background from thermometers positioned at 0\,m and $\pm$5.5\,m) are plotted in Fig.~\ref{fig:ndis}(b, c and d) 
to show the position of the selected cuts. 
Similarly, Fig.~\ref{fig:nndis} illustrates the selection criteria for the $nn$ disappearance search.
To remove the background from muon induced neutrons and after-muon detector activity, 
a 2\,ms veto of the entire detector volume following a muon traversal is applied for the case of 
the 3$^{rd}$ hit. The $\beta$-neutron emitter $^9$Li is produced by muon 
spallation and can give prompt and delayed signals. This background is 
suppressed by a 2\,s veto within a 3\,m radius cylinder around a well-reconstructed 
muon track. The entire volume is vetoed for 2\,s following energy depositions with more than 10$^6$ 
photo-electrons above what is expected from a minimum ionizing particle or if 
the muon is not well reconstructed. For $n$ disappearance the accidental background is estimated 
with a 150\,s to 1000\,s delayed-coincidence off-time window for the 3$^{rd}$ hit.  In the
$nn$ disappearance case a 10\,s to 1000\,s window is used. The number of expected 
accidental background events in the entire data set for the $n$ disappearance 
search is 0.82$\pm$0.26. For the $nn$ disappearance search, the significantly 
greater energy of the 3$^{rd}$ hit (see Fig.~\ref{fig:nndis}d) allows selection of 
the energy threshold cut to be larger than the energy of the main source of 
background; the number of expected accidentals is 0.018$\pm$0.010 events.

\begin{figure}[t]
\includegraphics[scale=0.44]{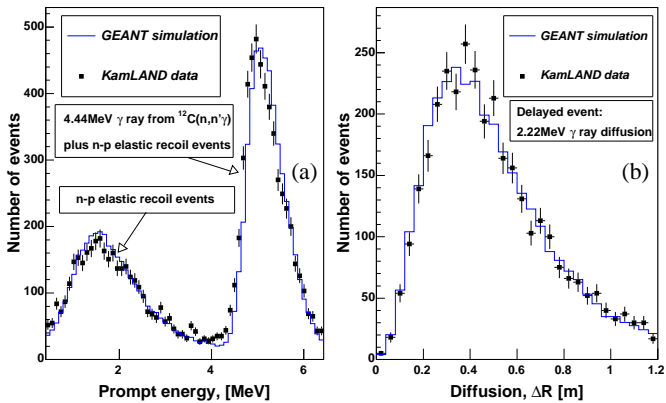}
\put(-143,85){(a)}
\put(-18,85){(b)}
\caption{KamLAND Am-Be data compared to the GEANT detector simulation: (a) prompt energy signal;
(b) diffusion of the 2.22\,MeV delayed $\gamma$ ray from neutron capture on proton}
\label{fig:ambe}
\end{figure}

The efficiency of the position and energy cuts is calculated using a Monte-Carlo 
procedure based on the GEANT code~\cite{Geant} set up to simulate the 
KamLAND detector response to the decay of the $^{11}$C$^*$ ($^{10}$C$^*$) 
nuclear system. The code includes new libraries \cite{GAMLIB} which give a more accurate 
description (compared to standard GEANT) of the $(n,\gamma)$ and $(n,n'\gamma)$ 
reactions.  
The decay schemes of $^{10}$C and $^8$B are well known, the 
$^{9}$C decay scheme is less well studied and the most recent 
data~\cite{Buchmann} are utilized for our simulations. The observed non-linear 
dependence of the energy estimation upon particle energy and type (for 
$\gamma$ rays and the neutron Am-Be source) is reproduced with a combination of a correction 
for scintillator light output quenching  \cite{Birks} and a theoretical correction for Cherenkov light emission for 
electrons and positrons (Fig.~\ref{fig:ambe}a). The energy and position are taken from the Monte Carlo simulation and
then smeared according to the resolution observed in calibration $\gamma$ ray and other data (Fig.~\ref{fig:ambe}b).
The reliability of neutron simulations in the LS was tested by comparison to simulations using the SCINFUL 
code~\cite{SCINFUL} and a good agreement was found. 
The calculated efficiencies for the de-excitation modes in Table~\ref{tab1} are given in 
Table~\ref{tab2}.

The systematic uncertainty of the predicted theoretical nuclear de-excitation 
branching ratios as estimated in \cite{Kolbe} is rather large, 
$\pm$30\%, while the experimental systematic uncertainties, such as the
fiducial volume uncertainty and the  uncertainty in neutron energy scale estimated by comparison
of the Am-Be data  and GEANT simulations (see also Table~\ref{tab2} and Ref.~\cite{Eguchi}), are 
significantly smaller. The total systematic uncertainty is $\sim$30\%.
The lifetime limits 
$\tau(n (nn) \rightarrow inv)$  for $n$ and $nn$ disappearance are 
calculated using  
\begin{eqnarray}
\tau (n (nn) \rightarrow inv) > \frac{N_0\,T}{x_{lim}} \sum_{i=1,2}{\epsilon_{\text{n(nn)}i} \,B_{\text{n(nn)}i}} ~,
\label{eq1}
\end{eqnarray}
where $N_0$ is the number of $s$-shell neutrons (or n pairs), $T$ is the detector livetime and $\epsilon_{\text{n(nn)}i}$
is the detection efficiency (given in Table~\ref{tab2}), $B_{\text{n(nn)}i}$ 
is the branching ratio from Table~\ref{tab1}, and 
${x_{lim}}$ is the largest number of events compatible with observation at a 
given Confidence Level. If classical confidence intervals for Poisson processes with
background (as described in Sec.~II-III of Ref.~\cite{Feldman}) are used, 
${x_{lim}}$\,=\,3.07\,(${x_{lim}}$\,=\,2.28) events at 90\%~CL, for 
neutron (two neutron) disappearance respectively.
The corresponding lifetime limits at 90\%~CL are $\tau (n \rightarrow inv) > 7.2 \times 10^{29} \text{y}$ and $\tau (nn \rightarrow inv) > 1.7 \times 10^{30} \text{y}$. 
If the Feldman-Cousins procedure \cite{Feldman} is used instead of the simple  
Poisson limit, ${x_{lim}}$\,=\,3.53\,events at 90\% CL (${x_{lim}}$\,=\,2.41\,events at 
90\% CL) and the corresponding lifetime limits are slightly reduced: 
$\tau~(n~\rightarrow~inv)~>~6.3~\times~10^{29}~\text{y}$ and $\tau~(nn~\rightarrow inv)~> 1.6~\times~10^{30}~\text{y}$. 

The effect of the uncertainty in the theoretical branching ratios and uncertainty for the number of expected 
background events is estimated using the program POLE \cite{Conrad}. 
Calculations using POLE are consistent with the results derived using the Feldman-Cousins procedure 
in the case of no systematic uncertainties and give ${x_{lim}}$\,=\,3.82~events at 90\% CL 
(${x_{lim}}$\,=\,2.75~events at 90\% CL) when the systematic uncertainties are included;
the lifetime limits are:
\begin{eqnarray}
\tau (n \rightarrow inv) > 5.8 \times 10^{29} \text{y at 90\% CL}  \\
\tau (nn \rightarrow inv) > 1.4 \times 10^{30} \text{y at 90\% CL} 
\end{eqnarray}

To summarize, about 838\, metric ton-years (1119\, metric ton-years) of KamLAND data have been 
analyzed in order to search for disappearance of a single neutron (a neutron 
pair) from the $\it{s}$~shell in $^{12}\text{C}$ which create subsequent nuclear 
de-excitations leading to three time and space correlated events. The 
observed number of events is consistent with that expected from the accidental 
coincidence background. This results in new improved limits on 
single neutron and two neutron invisible decay lifetimes. This KamLAND result is 
a factor of 3 better than the present best limit from the SNO collaboration~\cite{Ahmed} 
for $n$ disappearance, and more than 4 orders of magnitude better 
than the $nn$ disappearance limit set by the Borexino collaboration~\cite{Back}. 

The KamLAND experiment is supported by the COE program under grant 09CE2003 of 
the Japanese Ministry of Education, Culture, Sports, Science and Technology, and 
under the United States Department of Energy grant DEFG03-00ER41138. We are grateful to the Kamioka Mining 
and Smelting Company that provided service for activities in the mine.

\end{document}